\documentclass[a4paper]{mem}
\usepackage{natbib}
\usepackage{graphicx}
\usepackage[a4paper]{hyperref}
\def\msun{$M_\odot$}
\newcommand{\teff}{\mbox{$T_{\rm eff}$}}

\def\simgt{\lower.5ex\hbox{$\; \buildrel > \over \sim \;$}}
\def\simlt{\lower.5ex\hbox{$\; \buildrel < \over \sim \;$}}
\begin{document}

\title{Multiple helium abundances in Globular Clusters stars:
}

   \subtitle{Consequences for the Horizontal Branch and RR Lyrae}

\author{
Francesca D'Antona\inst{1}, Paolo Ventura\inst{1} \and Vittoria Caloi\inst{2} 
          }

  \offprints{F. D'Antona}

\institute{
Istituto Nazionale di Astrofisica --
Osservatorio Astronomico di Roma, Via di Frascati 33,
I-00040 Monteporzio, Italy
\and
INAF IASF, via del Fosso del Cavaliere, 00100 Roma
\email{dantona@oa-roma.inaf.it}
}

\authorrunning{D'Antona et  al. }

\titlerunning{Helium abundances in GCs}

\abstract{ Most inhomogeneities in the chemical composition of GC stars are due 
to primordial enrichment. The model today most credited is that the winds lost 
by high mass Asymptotic Giant Branch (AGB) stars, evolving during the first 
$\simlt$200Myr of the Clusters life, directly form a second generation of stars 
with abundance anomalies. The best indirect hint towards this suggestion is the 
recognition that some peculiarities in the Horizontal Branch (HB) stars 
distribution (blue tails, gaps, anomalous luminosity slope of the flat part of 
the HB) can be attributed to the larger helium abundance  in the matter, 
processed through Hot Bottom Burning, from which these stars are born. The 
model has been reinforced by finding a peculiar main sequence distribution in 
the cluster NGC 2808, which also has a bimodal HB distribution and an extended 
blue tail: the excess of  blue objects on the main sequence has been 
interpreted as stars with very high helium. We remark that the RR Lyr 
distribution may be affected by the helium spread, and this can be at the basis 
of the very long periods of the RRab variables of the metal rich clusters NGC 
6388 and NGC 6441,  longer than for the very metal poor Oosterhoff II clusters. 
These periods imply that the RR Lyr are brighter than expected for their 
metallicities, consistent with a larger helium abundance.

}
\maketitle{}

\section{Introduction}
The recent observations of abundance spreads among Globular Clusters stars,
now observed also at the turnoff (TO) and among the subgiants
\citep[e.g.,][]{gratton2001} show that these anomalies must be
attributed to some process of ``self--enrichment" occurring at the first
stages of the life of the cluster, during the epoch in which the Supernova
explosions were already finished (carrying easily away from the clusters
their high velocity ejecta) and the massive Asymptotic Giant Branch (AGB)
stars were evolving. At an epoch starting some $\sim 5 \times 10^7$yr from the
birth of the first stellar generation, the massive AGBs 
cycle their envelope material through hot CNO-cycle at the 
bottom of their convective envelopes (Hot Bottom Burning --HBB) and lose them  
in low velocity winds, which may
remain into the cluster, where they are either accreted on the already formed
stars \citep{dgc1983} or mixed with residual gas and give origin to a new
stellar generation \citep{cottrell-dacosta}. In the latest years we have 
suggested that the spreads in 
chemical abundances are actually due to the birth of successive generation of 
stars {\it directly} from the ejecta of the massive AGBs of the first 
generation. An important hint towards this self--enrichment model was
the interpretation of the morphology of 
extended HBs in some globular clusters in terms of a spread in the initial 
helium content of the cluster stars, and recently it received support
from the peculiar distribution of stars in the main sequence of NGC 2808 (see
later) and $\omega$Cen \citep{bedin2004, norris2004, piotto2005}. 
Here we put a note of warning on the interpretation of the RR Lyrae 
distribution in GCs, which may also be affected by the helium variations.

\section{The AGB models for Population II}

The most striking abundance anomaly in GC stars is the spread in Oxygen which
can reach a factor $\sim 10$\ in the intermediate metallicity clusters like
M13 and NGC 6752 \citep[e.g.][]{kraft1993}. This spread extends to the TO
and subgiant stars, as shown by \citet{gratton2001}, and therefore can not be
fully inputed to `in situ' mixing. \citet{ventura2001} found that, in low
metallicity stars, the process of `Hot Bottom
Burning' (HBB), that is the nuclear processing which occurs at the basis of
the convective envelope of massive AGBs, takes place at such large
temperatures ($\simeq 10^8$K) that the full CNO cycle operates and converts
Oxygen into Nitrogen. Therefore the envelopes of these stars have an Oxygen
abundance much smaller than the initial. The processing is more efficient in
the most massive AGBs, and progressively less efficient in the lower masses,
which have smaller temperatures at the basis of the convective envelope.
ON processing is possible only at low metallicities, and thus
for the HBB conditions typical of the massive AGBs in GCs. With this
knowledge, it is natural to attribute the spread in Oxygen of GC stars to
HBB and to some ``self--enrichment" mechanism from the envelopes
of AGB stars. The global quantitative scenario is still debated, 
and in particular, models predict very different results for the Oxygen--Sodium
anticorrelation found in many clusters 
\citep{denis2003,fenner2004, ventura-dantona2005}. 
Nevertheless, the computation of AGB
models is subject to severe uncertainties, due to the approximations made
both for convection and mass loss, so that the quantitative results must
be carefully explored, before we have a
clear and fully satisfactory scenario ---or before we reject it. In particular 
\cite{ventura-dantona2005} have shown that results differing by order of magnitudes
can be obtained for the HBB nucleosynthesis products, depending on the convection
model adopted. Figure 1 shows a summary of the problem: the model predictions
for the yields differ by orders of magnitude, 
if we compare the results by \cite{fenner2004} 
and by \cite{ventura-dantona2005}.

An interesting hint on the modalities of self--enrichment came in these years from
a totally different field of research: as remarked 
by \citet{ventura2001} and \citet{ventura2002}, the
models of AGB which show Oxygen depletion, also show a noticeable helium
enhancement: the helium content can be as high as Y=0.30 or more,
for the most massive AGB ejecta, although
starting from a mere Y=0.24 (the Big Bang abundance).
This result is particularly robust, as it is due primarily to the so called
`second dredge up' phase, which is much less model dependent than the third
dredge up associated with the thermal pulses.
   \begin{figure}
   \centering
   \includegraphics[width=6.5cm]{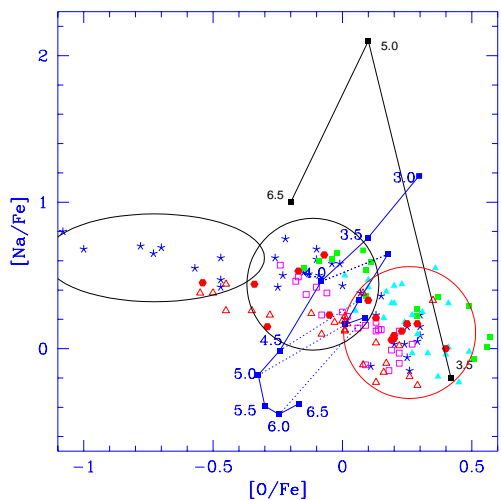}
      \caption{: O-Na anti-correlation for the stars of several globular clusters. 
	  NGC 2808 is labelled by full dots \citep{carretta2004}. 
	  The models by \cite{ventura-dantona2005} are labelled by masses from 
	  3 to 6.5 M$_\odot$, the models by Fenner et al. (2004) by dots of 3.5, 
	  5 and 6.5 M$_\odot$. The circle at the right includes the `normal' stars 
	  (first generation); the models should explain the abundances 
	  in the intermediate circle. 
	  The left circle stars, the most extreme, remain unexplained till now. }
	 \label{fig1}
   \end{figure}
If self-pollution is due to the matter lost from AGBs, the low mass
stars (M$\simlt$0.8\msun) presently evolving in GCs should be a
mixture of two populations, the first one, born together with the
intermediate mass population, and having the initial helium content, and a
second, additional, population more or less enriched in helium. The most
relevant feature is that, the larger is the helium content, the smaller is
the evolving mass for a given cluster age. 
E.g., the mass is reduced by $\sim 0.05$M$_\odot$\ for an increase 
in helium by 0.04. This mass difference is important 
for the \teff\ distribution on the HB, as first proposed by \cite{norris1981}.
In fact, if the same mechanism of mass loss operates on the ``standard
Y" and on the ``enhanced Y" stars along the giant branch and at the
helium flash, the final mass in HB will be several hundredths of solar mass
smaller, and therefore will have a {\it bluer} location. This has been
remarked by \citet{dantona2002}, who show that a
population of stars having enhanced Y {\it from the start} (that is, from the
main sequence) can explain the existence of extended blue tails in the HB of
some clusters, like NGC 6752 or M13, whose red giants show
the mentioned huge Oxygen spreads. In addition, the second star formation stage
may stop abruptly at some epoch, due, e.g., to
the presence of strong UV sources such as the planetary nebulae from
relatively low mass progenitors, leaving a
{\it discontinuity} between the helium content of the first generation (probably
the Big Bang abundance) and the {\it lowest} helium content of the second
generation. This produces a discontinuity in mass along the red giant
branch, which reflects in a discontinuity in mass along the HB. 
\cite{dantona2004} show that the
helium variation and discontinuity provide an interesting explanation for
the very peculiar distribution of stars in the
HB of NGC 2808, a conclusion which was reinforced by the discovery that the
main sequence of NGC 2808 presents an asymmetric color distribution which can 
best be explained by adding to the normal stars a population of 15--20\% of
stars with very high helium abundance (Y$\sim 40$\%). For
the helium distribution in the stars of
NGC~2808, see the extensive discussion in  \cite{dantona-bellazzini2005}.

\section{Are the HB luminosity and the RR Lyr period distribution
altered by the helium spread?}

The helium spread, although not altering in a significant way the
absolute luminosity of the RR Lyrae in clusters in which there is a
consistent ``first generation" population \citep{dantona2002} produces,
in the particular case of NGC 2808, the small but noticeable difference
in luminosity between the cool side of the blue HB and the hot side of
the red HB \citep{bedin2000}, which, so far, had not been consistently explained. 
There are other clusters showing a marked bimodality, such as
the metal rich ones NGC 6388 and NGC 6441 \citep{rich-sosin1997}. 
The metallicity of NGC 6441 has been recently confirmed to be very large
\citep{clementini2005} and is not consistent with the very long periods of the
RRab variables \citep{pritzl2000} in these two clusters. 
The marked slope of the horizontal part of the HB in both NGC 6388 and NGC 6441
(the bluer stars being more luminous),
can be attributed to the same self--enrichment mechanism which we
have described here: a fraction of the stars in these clusters
belongs to a ``second generation" with much larger helium content, and
the luminosity increases with \teff\ just because of the larger
helium abundance. This may also explain the long pulsation periods.
Thus {\it synthetic models of HB and RR Lyr stars distribution in GCs must take into
account the second parameter Y}.

Is it possible that there are clusters in which the first generation stars
have been completely lost, so that the helium content of {\it all} the stars
is larger than the primordial value? This has possibly occurred, e.g.,
in the classic ``second parameter" pair M3 and M13, according to \cite{caloi2005} 
suggestion. Can we get rid of the red HB stars of M3, in
order to obtain a fully blue HB like that of M13? We must invoke again two
stellar generations with different Y, but the mass function of the
first generation must peaked at intermediate mass stars, so that there are no
first generation low mass stars in M13.
The problem of a peculiar IMF for the first generation stars \citep{dantona2004}
receives some support from the observations of the Arches cluster 
\citep{stolte2005}.

\bibliographystyle{aa}

\end{document}